\documentclass{article}
\usepackage{spconf}
\usepackage{cite}
\usepackage{amsmath,amssymb,amsfonts}
\usepackage{algorithmic}
\usepackage{graphicx}
\usepackage{textcomp}
\usepackage{xcolor}
\usepackage{balance}
\usepackage{acronym}
\usepackage{hyperref}
\usepackage{csquotes}
\usepackage{adjustbox}
\usepackage{nicematrix}
\usepackage{booktabs}
\usepackage{multirow}
\usepackage{siunitx}
\usepackage{tikz,pgfplots}
\usepackage{cleveref}
\pgfplotsset{compat=newest}
\usepgfplotslibrary{colorbrewer}
\usepgfplotslibrary{fillbetween}
\usepgfplotslibrary{groupplots}

\makeatletter
\newcommand*{\org@overidelabel}{}
\let\org@overridelabel\AC@verridelabel
\renewcommand*{\AC@verridelabel}[1]{%
  \@bsphack
  \protected@write\@auxout{}{\string\AC@undonewlabel{#1@cref}}%
  \org@overridelabel{#1}%
  \@esphack
}%
\makeatother

\newcommand{\BE}{\begin{equation}\begin{aligned}}
\newcommand{\EE}{\end{aligned}\end{equation}}

\DeclareMathOperator{\cossim}{sim}

\DeclareMathOperator{\cost}{cost}
\DeclareMathOperator*{\argmax}{arg\,max}

\DeclareSIUnit{\decibel}{dB}

\acrodef{dtw}[DTW]{dynamic time warping}
\acrodef{hf}[HF]{high frequency}
\acrodef{itu}[ITU]{International Telecommunication Union}
\acrodef{kws}[KWS]{keyword spotting}
\acrodef{hfcc}[HFCC]{human factor cepstral coefficient}
\acrodef{mfcc}[MFCC]{mel-frequency cepstral coefficient}
\acrodef{snr}[SNR]{signal-to-noise-ratio}
\acrodef{awgn}[AWGN]{additive white Gaussian noise}
\acrodef{stft}[STFT]{short-time Fourier transform}

\begin{document}
\ninept
\title{Quantization-Based Score Calibration for Few-Shot Keyword Spotting with Dynamic Time Warping in Noisy Environments}

\name{Kevin Wilkinghoff$~^{1,2}$, Alessia Cornaggia-Urrigshardt$^{3}$, Zheng-Hua Tan$^{1,2}$}
\address{$^{1}$Department of Electronic Systems, Aalborg University, Denmark, $^{2}$Pioneer Centre for AI, Denmark\\ $^{3}$Fraunhofer FKIE, Wachtberg, Germany}

\maketitle

\begin{abstract}
Detecting occurrences of keywords with \ac{kws} systems requires thresholding
continuous detection scores.
Selecting appropriate thresholds is a non-trivial task, typically relying on
optimizing performance on a validation dataset.
However, such greedy threshold selection often leads to suboptimal performance
on unseen data, particularly in varying or noisy acoustic environments or
few-shot settings.
In this work, we investigate detection threshold estimation for template-based
open-set few-shot \ac{kws} using \acl{dtw} on noisy speech data.
To mitigate the performance degradation caused by suboptimal thresholds, we
propose a score calibration approach that operates at the embedding level by
quantizing learned representations and applying quantization error-based
normalization prior to DTW-based scoring and thresholding.
Experiments on \ac{kws}-DailyTalk with simulated \acl{hf} radio channels show that
the proposed calibration approach simplifies the selection of robust detection
thresholds and significantly improves the resulting performance.
\end{abstract}


\acresetall

\section{Introduction}
\Ac{kws} is the task of detecting spoken words or phrases, so-called keywords, in audio recordings.
Typical \ac{kws} applications are activating voice assistants \cite{michaely2017keyword}, searching for content in large databases \cite{moyal2013phonetic} or monitoring (radio) communication transmissions \cite{menon2017radio}.
Inherently, \ac{kws} is an open-set classification task as most spoken words or non-speech related sounds contained in the recordings do not correspond to any of the keywords of interest.
In addition, often only a few training samples are available for each keyword \cite{mazumder2021few-shot,rusci2023few-shot}, known as few-shot learning, and users are also interested in precise on- and offsets of detected keywords.
Furthermore, for many \ac{kws} applications only limited computational resources are locally available on small devices \cite{cioflan2024on-device}.
All these requirements make \ac{kws} a challenging task that is far from being solved.
This is especially true in noisy environments that emphasize many of the difficulties and negatively impact the performance \cite{lopez-espejo2021novel}.

\par

\Ac{kws} systems usually produce a temporal sequence of continuous detection scores for individual keyword classes.
This can be achieved using a sliding window approach or methods that inherently produce on- and offsets for detected events, such as \ac{dtw} \cite{sakoe1978dynamic}.
To turn a sequence of scores into a set of detections, the scores are binarized with detection thresholds and a suitable post-processing of the results is applied \cite{kim2019query}.
In general, estimating detection thresholds is non-trivial \cite{metze2014in-depth} and strongly benefits from well-calibrated scores \cite{karam2011towards}.
This is also true for state-of-the-art deep-learning based \ac{kws} systems \cite{lopez-espejo2022deep} that learn representations of audio segments suitable to discriminate between keywords \cite{ma2019hypersphere,mazumder2021few-shot,kim2022dummy}.
\par
Existing work on calibrating detection scores for \ac{kws} focuses on aligning the thresholds for different keyword classes \cite{karakos2013score,mamou2013system,pham2014discriminative,yuan2019verifying}, which is necessary for \ac{kws} systems that rely on models that are not trained to discriminate between target keywords.
However, for discriminatively trained state-of-the-art systems, this is less of a problem as the models are explicitly trained to output well-calibrated posterior probabilities.
Moreover, in open-set settings, where systems also need to predict on- and offsets of detected keywords, it is difficult to decide on which auxiliary scores to use for normalization, as different samples are aligned differently and thus will likely also have different on- and offsets.
In addition, the difficulty of estimating detection thresholds in noisy environments is not addressed.

\par

Although obtaining high-quality encodings of noisy speech requires sophisticated methods \cite{yang2021source-aware}, we propose to quantize learned representations of \ac{kws} systems with the aim of calibrating the detection scores and in turn improve \ac{kws} performance in noisy environments.
The intuition behind this proposal is that quantization can also be used for speech enhancement \cite{shaughnessy1988speech,zhao2022speech} and there is even evidence that quantization leads to more meaningful speech representations in general \cite{baevski2020wav2vec}.
Existing work on applying quantization to \ac{kws} is mostly focused on reducing the size of trained models \cite{mishchenko2019low-bit,peter2020resource-efficient,peter2022end-to-end,zeng2022sub}, leaving a knowledge gap that we aim to close.

\par

The contributions of this work are as follows.
First, we demonstrate that the performance of \ac{kws} in noisy environments using learned representations strongly depends on the quality of the estimated decision thresholds.
To address this, we propose a score calibration approach that involves quantizing learned speech representations and applying local density-based score normalization using quantization errors.
In few-shot open-set \ac{kws} experiments on simulated \ac{hf} radio communications, the proposed score calibration approach proves highly effective in simplifying the selection of robust detection thresholds and significantly improving the performance.

\section{Template-based KWS with DTW}
\Ac{dtw} \cite{sakoe1978dynamic} measures the similarity between two sequences of features in three steps:
First, pairwise similarities are computed between all features of one sequence to those of the other sequence.
Then, the costs are accumulated in a matrix by summing them up according to allowed step sizes and normalizing the accumulated costs at each position with the corresponding path length.
Finally, an optimal warping path with minimal accumulated costs is determined in this matrix using dynamic programming.
Throughout this work, we use the term \emph{similarity} to denote the scalar-valued cosine similarity or inner product between unit-normalized embeddings, while \emph{cost} refers to a non-negative scalar quantity used by \ac{dtw}.
For sub-sequence \ac{dtw} \cite{mueller2007information}, the start and end points of the warping paths for one of the sequences can be shifted, corresponding to on- and offsets of detected events.
In the context of \ac{kws}, this enables one to align a sequence of features from a query keyword sample with a partial feature sequence of an arbitrarily long test recording.
Alternatively, Fr\'{e}chet means, e.g. estimated with \ac{dtw} barycenter averaging \cite{petitjean2011global}, or multi-sample \ac{dtw} \cite{wilkinghoff2024multi} can be used to reduce the computational complexity.
\par
Traditionally, speech features such as \acp{mfcc} or \acp{hfcc} are used for \ac{kws} with \ac{dtw} \cite{von2010perceptual}. 
In contrast to these traditional frame-level acoustic features, we operate on learned frame-level embeddings obtained from a discriminatively trained embedding model based on the TACos loss \cite{wilkinghoff2024tacos}. 
This loss extends the AdaCos loss \cite{zhang2019adacos} from learning individual embeddings to sequences of embeddings with the same temporal resolution as the input spectrogram.
More concretely, spectrograms are divided into short overlapping segments, and the embedding model is trained to predict both the keyword class and the position of each segment within the original keyword sample.
This enables the model to learn discriminative embeddings that evolve over time and are therefore well suited to be used as templates for \ac{dtw}.
After training the model, each segment of the spectrogram
$x(i_\text{seg})\in\mathbb{R}^{T\times M}$
is converted to a sequence of embeddings
$\phi(x(i_\text{seg}))\in\mathbb{R}^{T\times D}$,
as illustrated in \Cref{fig:embeddings}.
After aggregating embeddings belonging to overlapping segments,
a spectrogram
$x_\text{sample}\in\mathbb{R}^{T_\text{sample}\times M}$
of arbitrary length is mapped to a frame-aligned sequence of embeddings
$\phi(x_\text{sample})\in\mathbb{R}^{T_\text{sample}\times D}$.
\iftrue
\begin{figure}
    \centering
    \vspace*{-0.45cm}
    \begin{adjustbox}{width=1.05\columnwidth}
          \hspace*{-0.1\columnwidth}
          \includegraphics{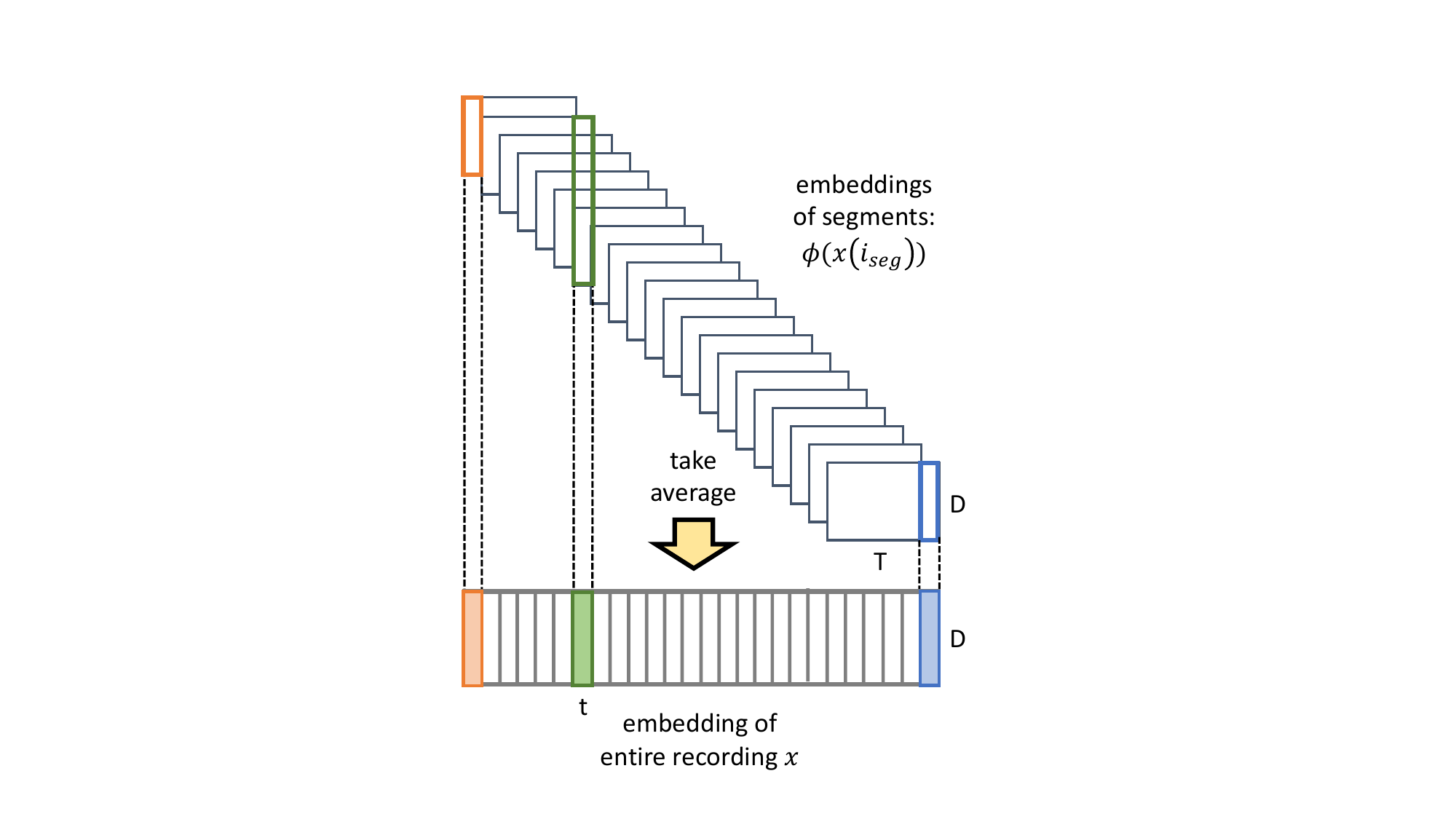}
    \end{adjustbox}
    \vspace*{-0.8cm}
    \caption{Illustration of combining the embeddings belonging to different segments of a single recording. Adapted from \cite{wilkinghoff2024audio}.}
    \label{fig:embeddings}
\end{figure}
\fi
\par
Let $N_\text{kw}$ and $N_\text{pos}$ denote the numbers of keyword and positional
classes of the TACos loss. For each keyword $i_\text{kw}$ and position
$i_\text{pos}$, let $\mathcal{C}_{i_\text{kw},i_\text{pos}}\subset\mathbb{R}^D$
denote the corresponding set of trainable centers, and let
$\cossim(\cdot,\cdot)$ denote the cosine similarity.
The embedding model
$\phi:\mathbb{R}^{T\times M}\rightarrow\mathbb{R}^{T\times D}$
is trained by computing the scalar similarity
\begin{equation}
\cossim_\text{sets}(\phi(x(i_\text{seg})),\mathcal{C}_{i_\text{kw},i_\text{pos}})
=
\frac{1}{T}\sum_{t=1}^{T}
\max_{c\in\mathcal{C}_{i_\text{kw},i_\text{pos}}}
\cossim(\phi(x(i_\text{seg}))_t,c),
\end{equation}
which is used as input to the AdaCos loss.

During inference with subsequence \ac{dtw}, the cost matrix between a query sample
$x_\text{query}$ and a test sample $x_\text{test}$ is defined as
\begin{equation}
\cost(x_\text{query},x_\text{test})_{i,j}
=
1-\langle\phi(x_\text{query})_i,\phi(x_\text{test})_j\rangle,
\end{equation}
where all embeddings are $\ell_2$-normalized.

\section{Quantization-based score calibration}
\begin{figure}
    \centering
    \vspace*{-0.15cm}
    \begin{adjustbox}{width=\columnwidth}
          \includegraphics{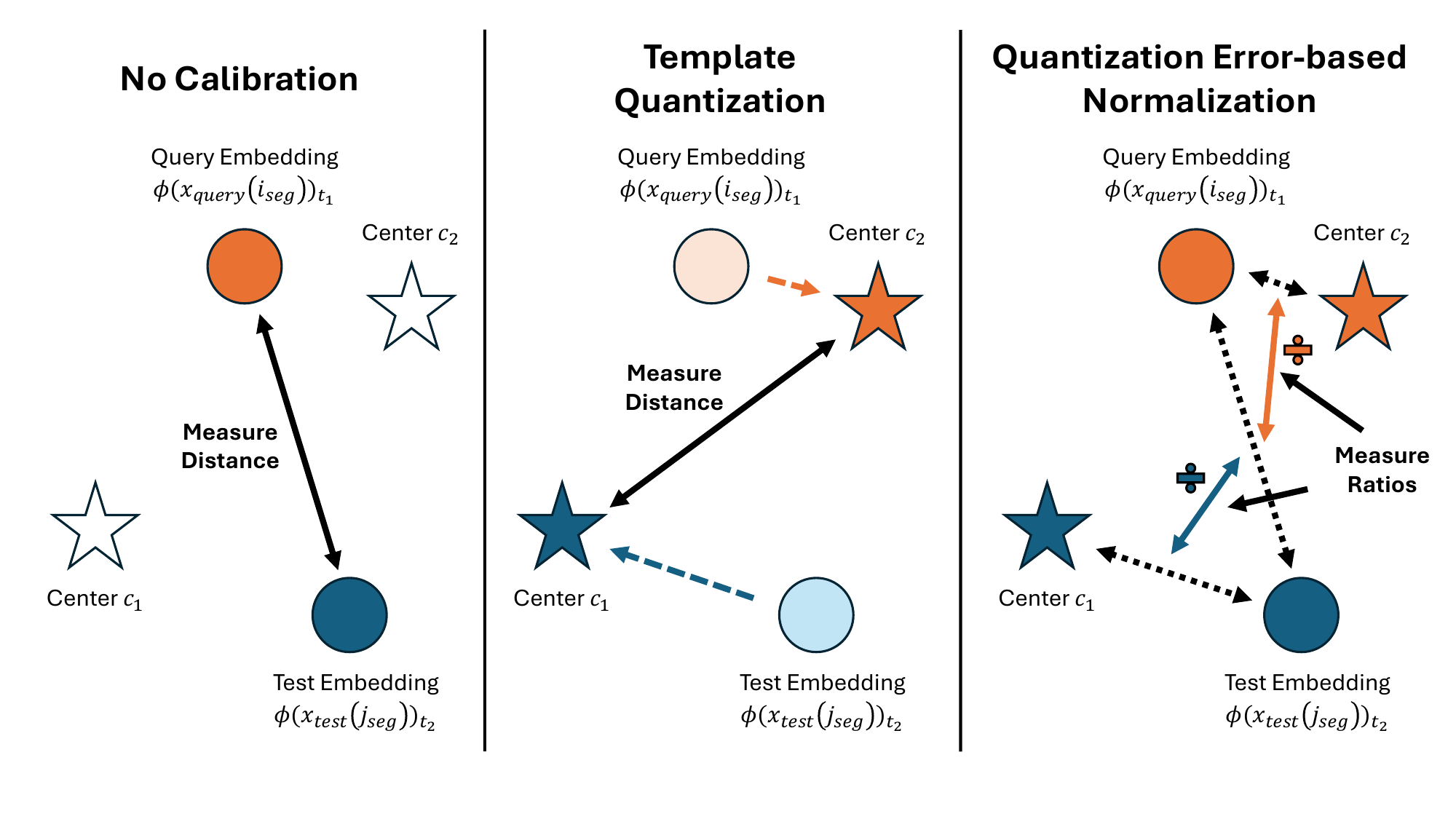}
    \end{adjustbox}
    \vspace*{-0.8cm}
    \caption{Illustration of the quantization steps. Without quantization, embeddings
    are directly compared (left). With quantization, the closest centers are compared
    (middle). When applying quantization error normalization, embedding similarities
    are adjusted relative to their quantization errors (right).}
    \label{fig:quantization}
\end{figure}

The proposed quantization-based score calibration approach combines two complementary
steps, each of which can also be applied individually:
1) quantizing embeddings and
2) normalizing embeddings using the quantization error.
The underlying idea is to reduce the impact of acoustic noise and the resulting
performance degradation.
Both steps utilize the centers learned during training of the embedding model and
are applied to individual embeddings prior to combining the embeddings belonging
to different segments.
\Cref{fig:quantization} illustrates the individual steps.

\subsection{Step 1: Template quantization}
The first step replaces each embedding by its closest center learned with the TACos
loss during training, defined as
\BE
\kappa(\phi(x(i_\text{seg}))_t)
:=
\argmax_{\substack{i_\text{kw}=1,\dots,N_\text{kw}\\
i_\text{pos}=1,\dots,N_\text{pos}\\
c\in \mathcal{C}_{i_\text{kw},i_\text{pos}}}}
\langle\phi(x(i_\text{seg}))_t,c\rangle
\in\mathbb{R}^{D},
\EE
for all $t=1,\dots,T$.
The underlying assumption is that the learned centers are less sensitive to
per-sample noise, as they capture information aggregated over many training samples.

\subsection{Step 2: Quantization error-based normalization}
Alternatively, embeddings can be scaled using their quantization error by setting
\BE
\nu(\phi(x(i_\text{seg}))_t)
:=
\frac{\phi(x(i_\text{seg}))_t}
{1+\max_{\substack{i_\text{kw}=1,\dots,N_\text{kw}\\
i_\text{pos}=1,\dots,N_\text{pos}\\
c\in \mathcal{C}_{i_\text{kw},i_\text{pos}}}}
\langle\phi(x(i_\text{seg}))_t,c\rangle}
\in\mathbb{R}^{D},
\EE
for all $t=1,\dots,T$.
Similarly to local density-based anomaly score normalization \cite{wilkinghoff2025keeping},
this scaling aims to reduce domain mismatch caused by different noise conditions.

\subsection{Combined score calibration approach}
To combine both steps, we sum the modified embeddings prior to aggregation:
\BE
\gamma(\phi(x(i_\text{seg}))_t)
:=
\kappa(\phi(x(i_\text{seg}))_t)+\nu(\phi(x(i_\text{seg}))_t)
\in\mathbb{R}^D.
\EE
Due to the linearity of the inner product, the contributions of both embedding-level
modifications add up in the resulting similarity scores used by \ac{dtw}.

\section{Experimental evaluation}

\subsection{Dataset}
For the experiments, we used the open-set few-shot \ac{kws} dataset \ac{kws}-Dailytalk \cite{wilkinghoff2024tacos} based on DailyTalk \cite{lee2023dailytalk}.
In contrast to other \ac{kws} datasets such as Speech Commands \cite{warden2018speech}, the goal is to not only predict the correct keyword but also to determine the on- and offsets of detected keywords, making the task more challenging.
\Ac{kws}-Dailytalk aims at detecting the following $15$ keywords in clean recordings taken from English conversations: \emph{afternoon}, \emph{airport}, \emph{cash}, \emph{credit card}, \emph{deposit}, \emph{dollar}, \emph{evening}, \emph{expensive}, \emph{house}, \emph{information}, \emph{money}, \emph{morning}, \emph{night}, \emph{visa} and \emph{yuan}.
The dataset is divided into a training set, a validation set and a test set.
The training set contains $5$ isolated samples of keywords (shots) per class and has a total duration of just $39$ seconds.
The validation and test sets consist of 156 and 157 sentences, respectively, containing several or none of the keywords.
Both sets are approximately $10$ minutes long and each keyword class appears roughly $12$ times in each set.
To measure \ac{kws} performance, the micro-averaged event-based F-score was used.
For all experiments, decision thresholds were chosen to maximize the F-Score on the validation set.
\par
Motivated by applications involving mission-critical communications, for which analog radio still plays an important role \cite{fritz2024analyzing}, an \ac{hf} radio channel simulation based on a Watterson model \cite{watterson1970} and \ac{awgn} was used to create noisy versions of \ac{kws}-DailyTalk.
Specifically, we applied a Watterson model for mid-latitude radio wave propagation under moderate conditions according to the \acs{itu} standard \cite{itur2000}, which corresponds to a multipath differential time delay of \SI{1}{\milli\second} and a Doppler spread of \SI{0.5}{\hertz}.
For the \ac{awgn}, \acp{snr} ranging from \SI{-12}{\decibel} to \SI{30}{\decibel} in steps of \SI{3}{\decibel} were used, resulting in $15$ versions of the dataset, one for each \ac{snr} value.
\iftrue
\subsection{Implementation details}
For all experiments, we used the TACos loss-based \ac{kws} system proposed in \cite{wilkinghoff2024tacos} with the following parameter choices.
For pre-processing, all waveforms were re-sampled to \SI{16}{\kilo\hertz}, high-pass filtered at \SI{50}{\hertz} and their amplitudes were normalized to $1$.
Then, log-mel spectrograms with $64$ mel-frequency bins were extracted using an \acs{stft} with Hanning-weighted windows of size $1024$ and a hop size of $256$.
The embedding model converts the log-mel spectrograms into sequences of embeddings and uses a modified ResNet architecture \cite{wilkinghoff2021twodimensional} with $713,486$ parameters when omitting the loss parameters.
This architecture consists of $4$ times two residual blocks, each consisting of convolutional layers with $3\times3$ filters, max-pooling along the frequency dimension and $20\%$ dropout \cite{srivastava2014dropout}.
After these blocks, a global max-pooling operation and a linear embedding layer with a dimension of $128$ are applied to the frequency and channel dimension, respectively.
For the AdaCos loss \cite{zhang2019adacos} used when training, the number of clusters per class was set to $N_C=16$.
The embedding model was trained for $1000$ epochs with a batch size of $32$ using Adam \cite{kingma2015adam}.
Apart from the classes corresponding to different keywords and different positions of speech segments within these keywords, an additional \emph{no-speech} class based on the background noise samples from SpeechCommands \cite{warden2018speech} and temporally reverted segments were used as negative samples belonging to none of the keywords or positions.
During training, random oversampling was applied to balance the classes and mixup \cite{zhang2017mixup} as well as SpecAugment \cite{park2019specaugment} were used for data augmentation.
As a backend, multi-sample \ac{dtw} \cite{wilkinghoff2024multi} with the step sizes $(1,1)$, $(2,1)$ and $(1,2)$ was used.
The results were post-processed by first shortening overlapping detections, retaining only the event with the highest score at each time step.
Subsequently, detections shorter than half of the length of the corresponding query template were discarded.
As an additional baseline system, a \ac{kws} system based on \acp{hfcc} was used, using a \SI{40}{\milli\second} window and a \SI{10}{\milli\second} step size with the same \ac{dtw} backend as used for the learned embeddings.
\fi

\subsection{Effectiveness of the proposed score calibration approach}
\begin{table}
    \centering
    \vspace{-6pt}
    \caption{F-scores obtained on \ac{kws}-DailyTalk for different \acsp{snr}. $95\%$ confidence intervals over five independent trials are shown. Decision thresholds maximize the F-score on the validation set.}
    \begin{adjustbox}{max width=\columnwidth}
    \begin{NiceTabular}{r|ccc|ccc}
        \toprule
        &\multicolumn{3}{c}{validation set}&\multicolumn{3}{c}{test set}\\
        \cmidrule(lr){2-4}\cmidrule(lr){5-7}
        &&\multicolumn{2}{c}{embeddings}&&\multicolumn{2}{c}{embeddings}\\
        \cmidrule(lr){3-4}\cmidrule(lr){6-7}
        \acs{snr} & \acs{hfcc} & no calibration & with calibration& \acs{hfcc} & no calibration & with calibration\\
        \midrule
        $\SI{-12}{\decibel}$ & $4.7$ & \pmb{$10.8\pm2.2$} & $10.3\pm2.4$& \pmb{$1.9$} & $0.4\pm1.2$ & $0.4\pm0.7$\\
        $\SI{-9}{\decibel}$ & $3.7$ & $13.1\pm1.8$ & \pmb{$14.3\pm1.4$}& $5.4$ & $5.2\pm5.0$ & \pmb{$9.7\pm2.8$}\\
        $\SI{-6}{\decibel}$ & $3.8$ & $21.6\pm2.6$ & \pmb{$24.0\pm2.3$} & $4.7$ & $11.7\pm4.3$ & \pmb{$13.8\pm1.5$}\\
        $\SI{-3}{\decibel}$ & $7.8$ & $25.0\pm4.2$ & \pmb{$25.3\pm2.8$} & $6.3$ & $14.1\pm3.3$ & \pmb{$15.7\pm2.0$} \\
        $\SI{0}{\decibel}$ & $8.8$ & $32.1\pm4.0$ & \pmb{$33.2\pm1.5$} & $10.4$ & $10.3\pm7.3$ & \pmb{$15.8\pm6.3$}\\
        $\SI{3}{\decibel}$ & $14.0$ & \pmb{$38.1\pm2.9$} & $37.8\pm3.4$ & $12.5$ & $13.1\pm11.1$ & \pmb{$27.0\pm7.2$} \\
        $\SI{6}{\decibel}$ & $24.0$ & \pmb{$50.7\pm2.6$} & $48.8\pm3.2$ & $26.8$ & $18.7\pm11.6$ & \pmb{$30.0\pm4.6$}\\
        $\SI{9}{\decibel}$ & $25.3$ & \pmb{$53.2\pm5.1$} & $50.3\pm4.2$ & $24.0$ & $22.9\pm4.4$ & \pmb{$32.1\pm3.6$} \\
        $\SI{12}{\decibel}$ & $31.4$ & \pmb{$58.8\pm4.0$} & $55.8\pm2.5$ & $33.1$ & $32.8\pm8.0$ & \pmb{$41.9\pm2.7$}\\
        $\SI{15}{\decibel}$ & $33.2$ & \pmb{$61.6\pm2.6$} & $60.5\pm3.4$ & $35.3$ & $34.2\pm10.5$ & \pmb{$42.3\pm4.4$} \\
        $\SI{18}{\decibel}$ & $35.9$ & \pmb{$59.1\pm3.6$} & $57.8\pm2.4$ & $36.5$ & $37.9\pm5.8$ & \pmb{$44.8\pm7.7$} \\
        $\SI{21}{\decibel}$ & $44.8$ & \pmb{$61.7\pm2.3$} & \pmb{$61.7\pm2.7$} & $44.4$ & $43.0\pm1.1$ & \pmb{$51.6\pm1.5$} \\
        $\SI{24}{\decibel}$ & $43.6$ & \pmb{$67.2\pm4.6$} & $63.6\pm1.7$ & $47.7$ & $50.4\pm4.9$ & \pmb{$53.8\pm5.3$} \\
        $\SI{27}{\decibel}$ & $48.4$ & \pmb{$69.7\pm3.3$} & $68.1\pm1.9$ & $44.7$ & \pmb{$61.6\pm4.7$} & $60.7\pm4.1$ \\
        $\SI{30}{\decibel}$ & $47.9$ & \pmb{$68.4\pm1.4$} & $67.9\pm2.0$ & $46.1$& \pmb{$62.3\pm1.7$} & $61.2\pm3.9$ \\
        \midrule
        Average & $25.2$ & \pmb{$46.1$} & $45.1$ & $25.3$ & $27.9$ & \pmb{$33.4$}\\
        \bottomrule
    \end{NiceTabular}
    \end{adjustbox}
    \label{tab:performance}
\end{table}
First, we verified the effectiveness of the proposed score calibration approach.
The results can be found in \Cref{tab:performance}, from which several observations can be made.
First, a lower \ac{snr} leads to worse performance for all methods, which is not surprising.
Furthermore, templates based on learned embeddings outperform templates based on \acp{hfcc} with a large margin when using an optimal decision threshold (as seen on the validation set) as well as under high \acp{snr}, i.e. of \SI{27}{\decibel} and \SI{30}{\decibel}, when using estimated decision thresholds.
This is consistent with the findings presented in \cite{wilkinghoff2024tacos,wilkinghoff2024multi}.
However, this is not the case when using estimated decision thresholds in noisy conditions, as shown in the comparison on the test set.
Without applying score calibration, the performance of the learned embeddings varies significantly and can even be worse than that of \acp{hfcc} (cf. the performance for an \ac{snr} of \SI{6}{\decibel}).
As we will show in \Cref{subsec:differences}, the main reason for this performance degradation is that the estimated decision thresholds are highly suboptimal.
Last but not least, the proposed calibration approach substantially improves the performance in noisy conditions on the test set over not calibrating the scores, while only having marginal performance degradations on the validation set and in less noisy conditions.
As a result, the embeddings consistently achieve a significantly higher performance than \acp{hfcc} when the proposed calibration approach is applied.

\subsection{Quality assessment of estimated decision thresholds}
\label{subsec:differences}
\begin{figure}
    \centering
    \begin{adjustbox}{max width=0.9\columnwidth}
          \begin{tikzpicture}
\begin{groupplot}[
    group style={
    group size=2 by 1,
    horizontal sep=2cm,},
	axis y line*=left,
    axis x line*=bottom,
    xmin=-12,
    xmax=30,
    xlabel=SNR \lbrack \si{\decibel}\rbrack,
    height=5.5cm,
    width=6cm,
    xticklabel style={align=center},
    yticklabel style={align=center},
    typeset ticklabels with strut,
    xlabel near ticks,
    ylabel near ticks,
    yticklabel style={xshift=-0.2cm},
    nodes near coords style={/pgf/number format/.cd,fixed zerofill,precision=2},
    ymajorgrids,
    scaled ticks = false,
    scaled y ticks = false,
    cycle list/Paired
]
\nextgroupplot[title=difference of thresholds, ylabel=$\theta_\text{opt}-\theta_\text{est}$, legend style={at={(1.2,1.35)},anchor=north,legend columns=3,/tikz/every even column/.append style={column sep=0.5cm}},ymin=-0.1,ymax=0.1,ytick={-0.1,-0.05,0,0.05,0.1},yticklabels={-0.1,-0.05,0,0.05,0.1}]
\legend{,no calibration (baseline),, with proposed calibration}
\addplot[name path=upper-without, fill=none, draw=none, forget plot] coordinates {(-12,0.006+0.024)(-9,-0.007+0.012)(-6,-0.01+0.013)(-3,-0.013+0.011)(-0,-0.025+0.024)(3,-0.043+0.014)(6,-0.031+0.032)(9,-0.033+0.012)(12,-0.022+0.012)(15,-0.031+0.016)(18,-0.027+0.01)(21,-0.034+0.012)(24,-0.025+0.012)(27,-0.008+0.013)(30,-0.017+0.007)};
\addplot[name path=lower-without, fill=none, draw=none, forget plot] coordinates {(-12,0.006-0.024)(-9,-0.007-0.012)(-6,-0.01-0.013)(-3,-0.013-0.011)(0,-0.025-0.024)(3,-0.043-0.014)(6,-0.031-0.032)(9,-0.033-0.012)(12,-0.022-0.012)(15,-0.031-0.016)(18,-0.027-0.01)(21,-0.034-0.012)(24,-0.025-0.012)(27,-0.008-0.013)(30,-0.017-0.007)};
\addplot+[opacity=0.5] fill between[of=lower-without and upper-without];
\addplot+[mark=triangle*,line width=1pt] coordinates {(-12,0.006)(-9,-0.007)(-6,-0.01)(-3,-0.013)(0,-0.025)(3,-0.043)(6,-0.031)(9,-0.033)(12,-0.022)(15,-0.031)(18,-0.027)(21,-0.034)(24,-0.025)(27,-0.008)(30,-0.017)};
\addplot[name path=upper-with, fill=none, draw=none, forget plot] coordinates {(-12,-0.018+0.039)(-9,-0.003+0.006)(-6,0.003+0.01)(-3,0+0.01)(0,0.017+0.02)(3,0.022+0.049)(6,0.009+0.019)(9,0.023+0.055)(12,0.012+0.032)(15,0.016+0.028)(18,0.007+0.014)(21,0.017+0.028)(24,0.003+0.009)(27,-0.001+0.004)(30,-0.001+0.005)};
\addplot[name path=lower-with, fill=none, draw=none, forget plot] coordinates {(-12,-0.018-0.039)(-9,-0.003-0.006)(-6,0.003-0.01)(-3,0-0.01)(0,0.017-0.02)(3,0.022-0.049)(6,0.009-0.019)(9,0.023-0.055)(12,0.012-0.032)(15,0.016-0.028)(18,0.007-0.014)(21,0.017-0.028)(24,0.003-0.009)(27,-0.001-0.004)(30,-0.001-0.005)};
\addplot+[opacity=0.5] fill between[of=lower-with and upper-with];
\addplot+[mark=square*,line width=1pt] coordinates {(-12,-0.018)(-9,-0.003)(-6,0.003)(-3,0)(0,0.017)(3,0.022)(6,0.009)(9,0.023)(12,0.012)(15,0.016)(18,0.007)(21,0.017)(24,0.003)(27,-0.001)(30,-0.001)};

\nextgroupplot[title=difference in performance, ylabel=F-score($\theta_\text{opt}$)$-$F-score($\theta_\text{est}$) (in percent),ymin=0,ymax=26,ytick={0,5,10,15,20,25},yticklabels={0,5,10,15,20,25}]
\addplot[name path=upper-without, fill=none, draw=none, forget plot] coordinates {(-12,1.3+2.8)(-9,1.6+2.1)(-6,3.+1.9)(-3,2.6+2.6)(0,8.6+8.8)(3,18.4+6.4)(6,14.+11.7)(9,11.9+5.5)(12,11.8+7.2)(15,13.5+9.5)(18,9.6+4.3)(21,13.9+6.2)(24,8.8+6.7)(27,3.2+1.7)(30,5.4+2.2)};
\addplot[name path=lower-without, fill=none, draw=none, forget plot] coordinates {(-12,1.3-2.8)(-9,1.6-2.1)(-6,3.-1.9)(-3,2.6-2.6)(0,8.6-8.8)(3,18.4-6.4)(6,14.-11.7)(9,11.9-5.5)(12,11.8-7.2)(15,13.5-9.5)(18,9.6-4.3)(21,13.9-6.2)(24,8.8-6.7)(27,3.2-1.7)(30,5.4-2.2)};
\addplot+[opacity=0.5] fill between[of=lower-without and upper-without];
\addplot+[mark=triangle*,line width=1pt] coordinates {(-12,1.3)(-9,1.6)(-6,3.)(-3,2.6)(0,8.6)(3,18.4)(6,14.)(9,11.9)(12,11.8)(15,13.5)(18,9.6)(21,13.9)(24,8.8)(27,3.2)(30,5.4)};
\addplot[name path=upper-with, fill=none, draw=none, forget plot] coordinates {(-12,0.4+0.5)(-9,0.2+0.4)(-6,0.2+0.3)(-3,0.3+0.4)(0,0.9+0.9)(3,1.2+1.8)(6,0.9+1.6)(9,0.9+2.2)(12,0.3+0.7)(15,0.9+0.9)(18,0.5+0.8)(21,0.7+0.8)(24,0.2+0.4)(27,0.1+0.1)(30,0.2+0.2)};
\addplot[name path=lower-with, fill=none, draw=none, forget plot] coordinates {(-12,0.4-0.5)(-9,0.2-0.4)(-6,0.2-0.3)(-3,0.3-0.4)(0,0.9-0.9)(3,1.2-1.8)(6,0.9-1.6)(9,0.9-2.2)(12,0.3-0.7)(15,0.9-0.9)(18,0.5-0.8)(21,0.7-0.8)(24,0.2-0.4)(27,0.1-0.1)(30,0.2-0.2)};
\addplot+[opacity=0.5] fill between[of=lower-with and upper-with];
\addplot+[mark=square*,line width=1pt] coordinates {(-12,0.4)(-9,0.2)(-6,0.2)(-3,0.3)(0,0.9)(3,1.2)(6,0.9)(9,0.9)(12,0.3)(15,0.9)(18,0.5)(21,0.7)(24,0.2)(27,0.1)(30,0.2)};
\end{groupplot}
\end{tikzpicture}
    \end{adjustbox}
    \caption{Difference between the optimal and estimated thresholds (left) and between the performances obtained with these thresholds (right) on the test set of \ac{kws}-DailyTalk for different \acsp{snr} when calibrating and not calibrating the scores. $95\%$ confidence intervals over five independent trials are shown.}
    \label{fig:differences}
\end{figure}

As a second experiment, we verified whether the differences in performance with and without calibration were actually caused by the choice of decision thresholds.
To this end, we determined the differences between the decision thresholds that maximize the performance on the validation set and the oracle thresholds that maximize the performance on the test set, as well as the corresponding differences in test set performance, both with and without the proposed score calibration approach.
The results are depicted in \Cref{fig:differences} and the following observations can be made.
Without applying the score calibration, relatively small differences between the estimated and optimal thresholds lead to large differences in performance.
This shows that the \ac{kws} system is highly sensitive to the threshold used and that it is difficult to estimate a good decision threshold.
In contrast, when using the proposed score calibration approach,  the difference between the thresholds is slightly smaller but of similar magnitude as without score calibration.
However, the difference in performance is consistently small.
Therefore, the \ac{kws} system is not sensitive to the decision threshold after calibrating the scores and estimating a good decision threshold is relatively easy.

\subsection{Effect of individual steps}
As a third experiment, we examined the effect of both sub-steps of the proposed score calibration approach individually.
The results in \Cref{tab:ablation} show that only quantizing the embeddings (step 1) leads to a significantly higher performance on the test set in noisy conditions than step 2.
In contrast, applying quantization error-based normalization (step 2) leads to significantly higher performance under less noisy conditions (\ac{snr} greater than \SI{24}{\decibel}) and on the validation set than step 1.
Therefore, quantizing the embeddings mainly contributes to obtaining more noise-robust scores that can be thresholded more effectively, whereas normalizing the scores reduces the mismatch between the decision thresholds of different keywords.
However, since the embedding model is trained discriminatively, compared to not calibrating the scores, this performance improvement is only marginal (cf. performance shown in \Cref{tab:performance}).
Last but not least, the proposed calibration approach outperforms both of its individual sub-steps, showing that both steps are complementary.

\section{Limitations and future work}
Despite the improvements presented in this work, there is still a performance difference between the estimated and optimal decision thresholds, and more work is needed to close this gap.
One possibility that may improve the resulting performance is to simulate different noise conditions and perform quantization during training of the embedding model.
However, this requires access to clean training samples, which are usually not available in practical applications.
Another limitation is that within this work, threshold values are estimated for specific \acp{snr}.
However, in real-world applications collected signals usually have different \acp{snr} that are a-priori unknown and estimating the \ac{snr} is highly non-trivial.
Therefore, a single estimated decision threshold should perform well in all possible noise conditions without further adjustment.
Ideally, one does not even need to estimate a decision threshold by optimizing the performance on a validation set.
Last but not least, additional experiments with other noise conditions and with other datasets or embeddings-based \ac{kws} systems can be conducted to strengthen the findings.

\section{Conclusion}
\begin{table}
    \centering
    \vspace{-6pt}
    \caption{F-scores obtained on \ac{kws}-DailyTalk for different \acsp{snr}. $95\%$ confidence intervals over five independent trials are shown. Decision thresholds maximize the F-score on the validation set.}
    \begin{adjustbox}{max width=\columnwidth}
    \begin{NiceTabular}{r|ccc|ccc}
        \toprule
        &\multicolumn{3}{c}{validation set}&\multicolumn{3}{c}{test set}\\
        \cmidrule(lr){2-4}\cmidrule(lr){5-7}
        \acs{snr} & both steps & step 1 only & step 2 only & both steps & step 1 only & step 2 only\\
        \midrule
        $\SI{-12}{\decibel}$ & $10.3\pm2.4$ & $8.5\pm1.5$ & \pmb{$11.2\pm3.2$} & $0.4\pm0.7$ & \pmb{$0.9\pm0.9$} & $0.0\pm0.0$\\
        $\SI{-9}{\decibel}$ & $14.3\pm1.4$ & $12.7\pm2.1$ & \pmb{$14.6\pm2.2$} & \pmb{$9.7\pm2.8$} & $7.9\pm3.6$ & $4.3\pm5.4$\\
        $\SI{-6}{\decibel}$ & \pmb{$24.0\pm2.3$} & $19.6\pm2.0$ & $22.7\pm2.7$ & \pmb{$13.8\pm1.5$} & $12.4\pm3.7$ & $10.8\pm5.3$\\
        $\SI{-3}{\decibel}$ & $25.3\pm2.8$ & $21.7\pm1.7$ & \pmb{$25.5\pm4.1$} & $15.7\pm2.0$ & $13.4\pm2.5$ & \pmb{$16.4\pm3.9$}\\
        $\SI{0}{\decibel}$ & $33.2\pm1.5$ & $26.6 \pm1.2$ & \pmb{$33.7 \pm4.5$} & \pmb{$15.8\pm6.3$} & $15.5\pm8.1$ & $9.8\pm5.2$\\
        $\SI{3}{\decibel}$ & $37.8\pm3.4$ & $35.0\pm4.7$ & \pmb{$39.2\pm3.0$} & \pmb{$27.0\pm7.2$} & $26.6\pm7.7$ & $13.0\pm9.8$\\
        $\SI{6}{\decibel}$ & $48.8\pm3.2$ & $40.9\pm2.2$ & \pmb{$51.1\pm3.6$} & \pmb{$30.0\pm4.6$} & $29.9\pm5.1$ & $18.3\pm12.4$\\
        $\SI{9}{\decibel}$ & $50.3\pm4.2$ & $45.2\pm3.4$ & \pmb{$53.2\pm3.4$} & \pmb{$32.1\pm3.6$} & $30.9\pm7.3$ & $26.6\pm5.1$\\
        $\SI{12}{\decibel}$ & $55.8\pm2.5$ & $49.9\pm1.3$ & \pmb{$60.6\pm4.2$} & \pmb{$41.9\pm2.7$} & $36.8\pm2.9$ & $33.8\pm9.0$\\
        $\SI{15}{\decibel}$ & $60.5\pm3.4$ & $53.5\pm3.4$ & \pmb{$63.4\pm2.8$} & \pmb{$42.3\pm4.4$} & $41.4\pm3.7$ & $35.6\pm6.9$\\
        $\SI{18}{\decibel}$ & $57.8\pm2.4$ & $52.3\pm1.9$ & \pmb{$60.8\pm3.7$} & \pmb{$44.8\pm7.7$} & $41.8\pm5.8$ & $41.1\pm3.3$\\
        $\SI{21}{\decibel}$ & $61.7\pm2.7$ & $56.4\pm2.6$ & \pmb{$62.4\pm3.5$} & \pmb{$51.6\pm1.5$} & $51.2\pm4.1$ & $46.1\pm2.4$\\
        $\SI{24}{\decibel}$ & $63.6\pm1.7$ & $58.3\pm2.1$ & \pmb{$68.3\pm4.0$} & \pmb{$53.8\pm5.3$} & $48.8\pm3.0$ & $53.4\pm5.9$\\
        $\SI{27}{\decibel}$ & $68.1\pm1.9$ & $63.5\pm3.9$ & \pmb{$70.4\pm3.5$} & $60.7\pm4.1$ & $56.9\pm3.0$ & \pmb{$63.6\pm3.7$}\\
        $\SI{30}{\decibel}$ & $65.1\pm2.1$ & $62.2\pm1.7$ & \pmb{$67.9\pm2.0$} & $61.2\pm3.9$ & $55.5\pm3.2$ & \pmb{$64.0\pm3.7$}\\
        \midrule
        Average & $45.1$ & $40.4$ & \pmb{$47.0$} & \pmb{$33.4$} & $31.3$ & $29.1$\\
        \bottomrule
    \end{NiceTabular}
    \end{adjustbox}
    \label{tab:ablation}
\end{table}
In this work, a score calibration approach for template-based few-shot \ac{kws} with \ac{dtw} was proposed.
The approach consists of two steps that are applied prior to generating the templates: 1) quantizing embeddings and 2) normalizing scores based on the quantization error.
Experimental evaluations conducted under \ac{hf}-specific noise conditions simulated for \ac{kws}-DailyTalk revealed that without calibrating the scores the estimated decision thresholds are highly non-optimal causing the performance to degrade significantly.
Furthermore, it was shown that the proposed score calibration approach is effective in minimizing this performance degradation, and both steps of the approach are important components.

\bibliographystyle{IEEEbib-abbrev}
\bibliography{mybib}

\end{document}